\documentclass[twocolumn]{aastex701}

\usepackage{xcolor}
\usepackage{amsmath, graphicx}
\usepackage{array}
\usepackage{xspace}
\usepackage{tabularx}
\usepackage{makecell}

\newcommand{\Msun}{M_{\odot}}

\newcommand{\cpeak}{\textsc{Compactness Peaks}\xspace}
\newcommand{\cpeaknew}{\textsc{Compactness Peaks + Channels}\xspace}
\newcommand{\bpltwopeak}{\textsc{Broken Power Law + 2 Peaks}\xspace}
\newcommand{\bspline}{\textsc{B-Spline}\xspace}

\newcommand{\chieff}{\chi_{\rm eff}}
\newcommand{\chip}{\chi_{\rm p}}
\newcommand{\IsoL}{{\rm Iso\,1G\,BH_{L}}}
\newcommand{\IsoH}{{\rm Iso\,1G\,BH_{H}}}
\newcommand{\DynL}{{\rm Dyn\,1G\,BH_{L}}}
\newcommand{\DynH}{{\rm Dyn\,1G\,BH_{H}}}
\newcommand{\DynOneG}{{\rm Dyn\,1G}}
\newcommand{\DynTwoG}{{\rm Dyn\,2G}}

\newcommand{\BHL}{{\rm BH_{L}}}
\newcommand{\BHH}{{\rm BH_{H}}}

\newcommand{\FracDynTwoG}{\ensuremath{0.02^{+0.03}_{-0.01}}}

\newcommand{\FracIsoLp}{\ensuremath{0.66^{+0.10}_{-0.14}}}
\newcommand{\FracIsoHp}{\ensuremath{0.09^{+0.10}_{-0.07}}}
\newcommand{\FracDynLp}{\ensuremath{0.13^{+0.16}_{-0.10}}}
\newcommand{\FracDynHp}{\ensuremath{0.09^{+0.05}_{-0.04}}}

\newcommand{\GapWidth}{\ensuremath{3.5^{+6.1}_{-5.5}}}

\newcommand{\MmaxIsoLp}{\ensuremath{12.3^{+2.4}_{-1.3}}}

\newcommand{\MuIsoLp}{\ensuremath{10.2^{+1.0}_{-0.7}}}

\newcommand{\MuDynHp}{\ensuremath{31.9^{+2.7}_{-4.7}}}

\newcommand{\CostTruncIso}{\ensuremath{-0.68^{+0.27}_{-0.27}}}
\newcommand{\MuChiIso}{\ensuremath{0.15^{+0.09}_{-0.11}}}
\newcommand{\MuChiDyn}{\ensuremath{0.19^{+0.14}_{-0.16}}}
\newcommand{\MuChiDynTwoG}{\ensuremath{0.65^{+0.30}_{-0.22}}}

\newcommand{\PostMOneMaxIsoHp}{\ensuremath{39.8^{+17.7}_{-13.6}}}

\newcommand{\PostMOneMaxDynHp}{\ensuremath{51.8^{+15.7}_{-11.6}}}

\newcommand{\PostQMuIsoLp}{\ensuremath{0.80^{+0.15}_{-0.12}}}

\newcommand{\PostQSigIsoLp}{\ensuremath{0.17^{+0.16}_{-0.08}}}

\newcommand{\PostQMuIsoHp}{\ensuremath{0.52^{+0.34}_{-0.31}}}

\newcommand{\PostQSigIsoHp}{\ensuremath{0.23^{+0.23}_{-0.16}}}

\newcommand{\PostQMuDynLp}{\ensuremath{0.62^{+0.30}_{-0.31}}}

\newcommand{\PostQSigDynLp}{\ensuremath{0.19^{+0.26}_{-0.12}}}

\newcommand{\PostQMuDynHp}{\ensuremath{0.89^{+0.09}_{-0.09}}}

\newcommand{\PostQSigDynHp}{\ensuremath{0.16^{+0.14}_{-0.08}}}

\newcommand{\PostQMuDynTwog}{\ensuremath{0.55^{+0.10}_{-0.13}}}

\newcommand{\PostQSigDynTwog}{\ensuremath{0.09^{+0.16}_{-0.04}}}

\newcommand{\PostChiMuDynTwog}{\ensuremath{0.65^{+0.30}_{-0.22}}}

\newcommand{\PostLamb}{\ensuremath{2.90^{+0.81}_{-0.83}}}

\shorttitle{Compactness Peaks and Subpopulations}

\begin{document}

\title{Compactness Peaks and Subpopulations:\\ Probing Stellar Physics and Formation Channels of Merging Binary Black Holes}

\author[orcid=0000-0002-1819-0215, gname=Shanika, sname=Galaudage]{Shanika Galaudage}
\affiliation{Center for Interdisciplinary Exploration and Research in Astrophysics (CIERA), Northwestern University, 1800 Sherman Ave, Evanston, IL 60201, USA}
\affiliation{The Adler Planetarium, 1300 South DuSable Lake Shore Drive, Chicago, IL 60605, USA}
\email[show]{shanika.galaudage@northwestern.edu}
\correspondingauthor{Shanika Galaudage}

\begin{abstract}
The growing catalog of gravitational-wave detections from the LIGO-Virgo-KAGRA (LVK) collaboration reveals non-trivial structure in the binary black hole (BBH) mass distribution, including peaks near $m_1 \approx 10$ $\Msun$ and $m_1 \approx 35$ $\Msun$, a high-mass suppression consistent with the pair-instability supernova gap, and a possible dearth of systems near $\mathcal{M}\approx 10$--$12$ $\Msun$ in the chirp mass distribution that may map to a $\sim 10$--$15$ $\Msun$ component-mass gap arising from non-monotonic stellar core compactness.
We apply \textsc{Compactness Peaks + Channels}, a stripped-star-motivated five-component population model, to 152 binary black hole (BBH) mergers from GWTC-4.0, and find decisive preference ($\log_{10}\mathcal{B} = 7.69$) over the LVK \textsc{Broken Power Law + 2 Peaks} baseline.
The model decomposes the population into isolated first-generation (1G), dynamical 1G, and hierarchical second-generation (2G) channels with inferred fractions $0.75^{+0.11}_{-0.16}$, $0.22^{+0.16}_{-0.11}$, and $0.02^{+0.03}_{-0.01}$ respectively.
The $\sim 10\,M_\odot$ feature is sharply localized by the low-mass isolated component, with properties---narrow mass, near-equal mass ratios, and low partially-aligned spins---consistent with stripped-star binary evolution.
The $\sim 35\,M_\odot$ feature is primarily captured by the high-mass dynamical 1G component.
The hierarchical component is internally consistent with 2G+1G mergers, with elevated primary spins ($\mu_\chi^{\mathrm{Dyn\,2G}} = 0.65^{+0.30}_{-0.22}$, agreeing with the numerical-relativity prediction of $\sim 0.7$) and tightly constrained asymmetric mass ratios.
We localize the edges of the predicted compactness peaks (Iso\,1G\,BH$_\mathrm{L}$ turn-off at $12.3^{+2.4}_{-1.3}\,M_\odot$, Iso\,1G\,BH$_\mathrm{H}$ turn-on at $16.1^{+5.7}_{-5.3}\,M_\odot$), consistent with, but not yet requiring, a compactness-driven dearth between them.
We also find mild support for additional structure near $\sim20\,\Msun$, though its astrophysical origin remains unclear.
Our results support a multi-component description of the merging BBH population and motivate further tests of compactness-driven isolated evolution against alternative astrophysical scenarios.
\end{abstract}



\section{Introduction}
\label{sec:intro}

The detection of gravitational waves from merging compact binaries by the LIGO-Virgo-KAGRA (LVK) Collaboration \citep{AdLIGO,AdVirgo,KAGRA} has opened a new observational window onto the life cycles of massive stars and the dense stellar environments in which compact objects interact.
With over 150 binary black hole (BBH) mergers reported in the fourth gravitational-wave transient catalog \citep[GWTC-4.0;][]{GWTC4}, population studies are beginning to resolve structure in the mass, spin, and redshift distributions of merging binary black holes \citep[e.g.,][]{GWTC4pop,Callister:2023tgi}.
In the primary mass distribution, excess support is observed near $m_1 \approx 10\,\Msun$ and $m_1 \approx 35\,\Msun$ \citep[e.g.][]{GWTC4pop,Callister:2023tgi,Ray:2024hos,Sridhar:2025kvi}.

We also observe structure in chirp mass, $\mathcal{M} = (m_1 m_2)^{3/5}/(m_1 + m_2)^{1/5}$, which is the combination of component masses most precisely measured from the gravitational-wave signal \citep{Finn:1992xs,Poisson:1995ef}.
The chirp mass distribution shows statistically significant peaks near $\mathcal{M} \approx 8\,\Msun$, $14\,\Msun$, and $27\,\Msun$, together with a relative dearth of systems in the range $\mathcal{M}\approx10$--$12\,\Msun$ \citep[e.g.][]{Tiwari:2020otp,Tiwari:2023xff,Tiwari:2025lit,Willcox:2025poh}, henceforth referred to as the ``chirp-mass gap''.

A physically motivated interpretation of the low-mass structure was proposed by \citet{Schneider:2023mxe}, who showed that isolated binary evolution of hydrogen-envelope-stripped stars can naturally produce a bimodal black-hole component-mass distribution through the non-monotonic compactness structure of stellar cores.
The compactness $\xi=M/R(M)$ is a useful predictor of explodability and black-hole formation \citep{OConnor:2010moj}, and exhibits structure driven by late burning phases and shell interactions \citep{Sukhbold:2013yca,Schneider:2020vvh,Schneider:2023mxe}.
In the stripped-star models of \citet{Schneider:2023mxe}, this structure gives rise to a low-mass black-hole population, $\BHL$, near $m\simeq9$--$10\,\Msun$, and a higher-mass population, $\BHH$, above $m\gtrsim15\,\Msun$, with suppressed black-hole formation in between.
We refer to this predicted component-mass dearth as the ``compactness gap''.
Because isolated binary evolution preferentially produces near-equal-mass pairings, mixed $\BHL+\BHH$ binaries are expected to be rare, causing the component-mass dearth to map onto a chirp-mass dearth near $\mathcal{M}\approx10$--$12\,\Msun$ \citep{Schneider:2023mxe}.
The observed low-mass BBH population has also been argued to contain a subpopulation consistent with isolated binary evolution, with distinct mass-ratio and spin properties near the $\sim10\,\Msun$ peak \citep[e.g.][]{Godfrey:2023oxb,GWTC4pop}.
A closely related possibility is that the low-mass peak is sharpened by a distinct failed-supernova channel, in which efficient direct collapse over a narrow progenitor range produces $\sim10\,\Msun$ black holes separated from the broader high-mass population by a suppressed rate at $m_1\simeq12$--$16\,\Msun$ \citep{Disberg:2023gel,Legred:2026oiz}.
Both interpretations are motivated by the non-monotonic explodability of massive stars and by core-collapse simulations showing that stellar-mass black holes can form through multiple pathways \citep[e.g.][]{Sukhbold:2015wba,Burrows:2024pur,Burrows:2024wqv}.
Previous population analyses found the observed component mass structure to be broadly consistent with this compactness-driven picture \citep{Adamcewicz:2024jkr,Galaudage:2024meo}, although neither study found clear evidence for a gap-like feature.

Beyond the mass spectrum alone, recent analyses increasingly suggest that the BBH population is better described as a superposition of subpopulations with different mass-ratio, spin, and redshift behavior \citep[e.g.][]{Banagiri:2025dmy,Tong:2025xir,Sridhar:2025kvi,Farah:2026jlc,Vijaykumar:2026zjy,Ray:2026uur,Plunkett:2026pxt}. In this interpretation, the $\sim10\,\Msun$ and $\sim35\,\Msun$ structures need not arise from the same formation pathway: the former is often associated with isolated binary evolution and low-mass stellar-collapse physics, while the latter may be enhanced by dynamical assembly, hierarchical growth, or other channel-dependent selection effects \citep{Godfrey:2023oxb,Sridhar:2025kvi,Ray:2026uur}.

In this work we focus on the ``compactness gap'' hypothesis and extend the \cpeak analysis from \cite{Galaudage:2024meo} to introduce a five-component mixture model (\cpeaknew). In this model, the BBH population is decomposed into five astrophysically motivated components whose mass, mass-ratio, and spin distributions are designed to capture isolated 1G, dynamical 1G, and hierarchical-merger contributions.

This manuscript is organized as follows.
Section~\ref{sec:method} describes the data, fiducial model, and targeted model reductions.
Section~\ref{sec:results} presents the model-comparison results and the inferred population structure.
Section~\ref{sec:discussion} discusses the astrophysical interpretation and caveats.

\section{Methods}
\label{sec:method}

In this section we describe the data, model design, and analysis framework used to infer the BBH population hyperparameters.
We first outline the event selection criteria and the data products used (\ref{sec:data}).
We then describe the fiducial five-component \cpeaknew model (\ref{sec:model}).
Finally, we define the targeted model reductions and variations used to identify which aspects of the fiducial parameterization are most strongly supported by the data (\ref{sec:modelreductions}).

\subsection{Data}
\label{sec:data}

We analyze BBH mergers up to and including the first part of the fourth observing run (O4a), forming the fourth Gravitational Wave Transient Catalog \citep[GWTC-4.0;][]{GWTC4}, with false alarm rates FAR $< 1\,\mathrm{yr}^{-1}$, consistent with the selection used in the LVK population analysis \citep{GWTC4pop}.
We use publicly released posterior samples and injection sets from \citet{GWTC4methods,GWTC4pe,Essick:2025zed}.
GW190814 is excluded owing to the uncertain nature of its secondary component, which is consistent with either a neutron star or a low-mass black hole \citep{GW190814}.
GW231123 is similarly excluded because its inferred near-extremal spins place it in a regime where waveform systematics and population-model assumptions require additional dedicated treatment \citep{GW231123}.
Including this event is therefore beyond the scope of the present analysis.
We additionally exclude GW241011 and GW241110 \citep{GW241011_GW241110} as they were detected outside the O4a observing period.
This gives a total sample of 152 BBH events for our hierarchical analysis.

\subsection{Fiducial model}
\label{sec:model}

We construct the \cpeaknew model around two assumptions.
\begin{enumerate}
    \item The isolated binary channel is expected to be dominated by black-hole progenitors that have been stripped of their hydrogen envelopes through binary interaction, hereafter binary-stripped stars (BSS) \citep[e.g.,][]{Mandel:2018hfr,vanSon:2022myr}.
    \item Dynamically assembled BBH mergers may contain a mixture of black holes formed from binary-stripped stars and non-stripped stars (NSS), because black holes formed through different stellar-evolution pathways can later be paired dynamically \citep[e.g.,][]{Rodriguez:2015oxa,Rodriguez:2016kxx,Mapelli:2021gyv}.
\end{enumerate}

Throughout this work, the ``Iso 1G'', ``Dyn 1G'', and ``Dyn 2G'' labels reflect the astrophysical motivation of each component, not a classification of individual events; their inferred mass, mass-ratio, and spin properties are compared to those formation scenarios rather than used to assign sources to channels.

\begin{figure}
    \centering
    \includegraphics[width=\linewidth]{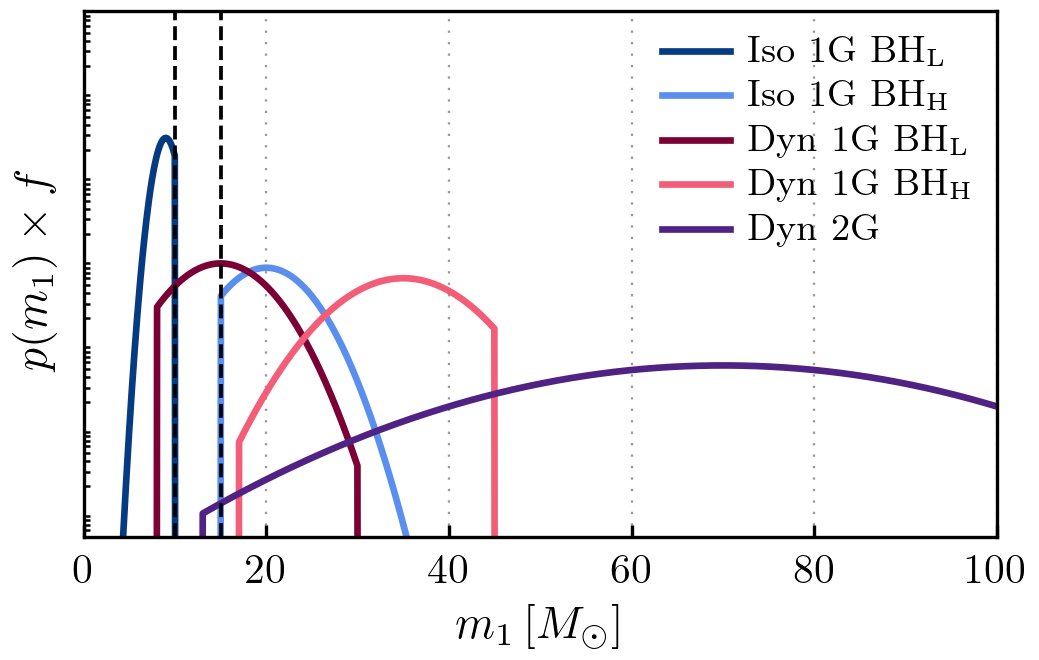}
    \caption{Illustration of the primary mass ($m_1$) components of the fiducial model \cpeaknew.
    The vertical dashed lines illustrate the predicted location of the ``compactness gap'' defined by the upper and lower edges of the $\IsoL$ and $\IsoH$ components respectively.
    The priors of the means, widths, and truncations of each peak are provided in Table~\ref{tab:compactnesspeaks_priors}.}
    \label{fig:schematic}
\end{figure}

The stripped-star calculations of \citet{Schneider:2023mxe} predict that compactness-driven structure should be most clearly visible in the BSS remnant population, whereas the corresponding structure is blurred for NSS because of the stronger metallicity dependence of their remnant masses.
We therefore allow the isolated channel to contain two compactness-motivated 1G components, $\IsoL$ and $\IsoH$, whose truncation bounds can produce a gap.
For the dynamical 1G population, we use two broader components, $\DynL$ and $\DynH$, which can overlap in mass and therefore need not preserve a clean compactness gap.
Finally, we include a fifth component, $\DynTwoG$, to capture a hierarchical-merger contribution.
Figure~\ref{fig:schematic} illustrates the five primary-mass components of the fiducial model.

Our fiducial \cpeaknew model has five components:
\begin{align}
\pi(m_1,q,\chi_{1,2},\cos&\theta_{1,2}, z\mid\Lambda) =\;\notag\\
&~f_{\IsoL}\,\pi^{\IsoL} \notag\\
&+f_{\IsoH}\,\pi^{\IsoH} \notag\\
&+f_{\DynL}\,\pi^{\DynL}\notag\\
&+f_{\DynH}\,\pi^{\DynH} \notag\\
&+f_{\DynTwoG}\,\pi^{\DynTwoG},
\label{eq:mixture}
\end{align}
where the mixture fractions satisfy $\sum_i f_i=1$ and are drawn from a symmetric Dirichlet prior.
Each component $\pi^X$ is modeled as a product of independent distributions in primary mass $m_1$, mass ratio $q$, spin magnitudes $\chi_1,\chi_2$, spin tilts $\cos\theta_1,\cos\theta_2$, and redshift $z$.
The component-specific choices for these distributions are as follows.
\begin{itemize}
    \item \textbf{Primary mass distributions ($m_1$):}
    All five components use truncated Gaussians in $m_1$ with component-specific means $\mu_{m_1}$, widths $\sigma_{m_1}$, and bounds $[m_{\min},m_{\max}]$.
    The isolated components, $\IsoL$ and $\IsoH$, are intended to capture the two compactness-motivated stripped-star black-hole populations.
    The dynamical 1G components, $\DynL$ and $\DynH$, are broader and are allowed to overlap in mass.
    The hierarchical component, $\DynTwoG$, uses a broad truncated Gaussian that can extend into the pair-instability supernova (PISN)-gap region \citep{Woosley:2016hmi,Farmer:2019jed,Gerosa:2021mno}.

    \item \textbf{Mass-ratio distributions ($q$):}
    Each component has a truncated Gaussian mass-ratio distribution on $q\in[0,1]$, with component-specific mean $\mu_q$ and width $\sigma_q$.

    \item \textbf{Spin-magnitude distributions ($\chi_{1,2}$):}
    Spin magnitudes are modeled as Gaussians truncated to $\chi\in[0,1]$.
    The two isolated 1G components share one spin-magnitude distribution, the two dynamical 1G components share another, and the hierarchical component uses an asymmetric prescription: $\chi_1$ is drawn from a separate 2G-motivated distribution, while $\chi_2$ is drawn from the dynamical 1G distribution.

    \item \textbf{Spin-orientation distributions ($\cos\theta_{1,2}$):}
    The isolated-channel spin tilts are modeled with a truncated Gaussian on $[\cos\theta_{\min},1]$, allowing preferential alignment but also permitting partial misalignment.
    The dynamical components have isotropic spin tilts by construction.

    \item \textbf{Redshift evolution ($z$):}
    All components share a single power-law redshift evolution,
    \begin{equation}
        \pi(z\mid\kappa) \propto \frac{dV_c}{dz}(1+z)^{\kappa-1},
    \end{equation}
    where $\kappa$ is the slope of the power law.
\end{itemize}
Details of the hyperparameters and prior ranges are provided in Appendix~\ref{app:priors}.

We perform hierarchical Bayesian inference to measure the population hyperparameters from the LVK gravitational-wave data, following the ``recycling'' framework described in Section~5 of \citet{Thrane:2018qnx}; see also \citet{Loredo:2004nn,Mandel:2018mve,Vitale:2020aaz}.
Our analysis uses \texttt{GWPopulation} \citep{gwpopulation} for the hierarchical inference, with posterior samples obtained using the nested sampler \texttt{dynesty} \citep{dynesty}.
To ensure that our Monte Carlo estimators of the population likelihood are converged, we discard hyperparameter samples whose log-likelihood variance exceeds $\sigma^2_{\ln \hat{\mathcal{L}}} = 1$ \citep{Talbot:2023pex}, consistent with the criterion adopted by \citet{GWTC4pop}.
We compare the fiducial model and its variants to the default LVK model, the \bpltwopeak baseline \citep{GWTC4pop}, using both the Bayes factor ($\mathcal{B}$) and the maximum likelihood ($\mathcal{L}_{\rm max}$).
All values are reported to the 90\% credible interval unless otherwise specified. 

\subsection{Model reductions}
\label{sec:modelreductions}

To identify which parts of the fiducial parameterization are most strongly supported by the data, we consider some targeted reductions of the models:
\begin{enumerate}
    \item [\textbf{D:}] Collapses the $\DynL$ and $\DynH$ components into a single dynamical 1G distribution;
    \item [\textbf{Q:}] Shares the mass-ratio distribution within channel type (one $q$ model for $\IsoL$ and $\IsoH$, and another for $\DynL$ and $\DynH$);
    \item [\textbf{S:}] Shares the spin-magnitude distribution between $\IsoL$, $\IsoH$, $\DynL$, and $\DynH$, collapsing all 1G components to have the same $\chi$ distribution.
    This reduction tests whether the isolated and dynamical 1G components prefer to have different distributions.
    \item [\textbf{T:}] Shares the spin-tilt distribution across all channels by using a mixture model between the aligned (Gaussian) and isotropic components where the mixing fraction is $\zeta$.
\end{enumerate}
We also consider further restrictions and analysis of feature requirements discussed in Appendix~\ref{app:modelcomp}. 


\section{Results}
\label{sec:results}

\begin{table*}
\centering
\begin{tabular}{l c c}
    \hline
    Model & $\log_{10}\mathcal{B}$ & $\Delta\log_{10}\mathcal{L}_{\rm max}$ \\
    \hline\hline
    \cpeaknew (fiducial) & $+0.00$ & $+0.00$ \\
    \cpeaknew [S] & $-0.56$ & $-0.89$ \\
    \cpeaknew [Q] & $-1.04$ & $-1.69$ \\
    \cpeaknew [T] & $-1.16$ & $-1.16$ \\
    \cpeaknew [D] & $-1.90$ & $-0.90$ \\
    \hline
    \textsc{Broken Power Law + 2 Peaks} & $-7.69$ & $-10.04$ \\
    \hline
\end{tabular}
\caption{
Model comparison relative to the fiducial \cpeaknew model.
Negative values indicate that a model is disfavored relative to the fiducial model.
[D] collapses the two dynamical 1G components into one; [Q] shares mass-ratio distributions within channel type; [S] shares the 1G spin-magnitude distribution between the isolated and dynamical 1G components; and [T] uses a single shared isotropic$+$aligned spin-tilt mixture across channels.
The LVK \bpltwopeak model is shown for reference.
}
\label{tab:bf}
\end{table*}

\subsection{Model comparison}
\label{sec:modelcomp}

We refer to \cite{Kass:1995loi} to quantify the significance of a result based on the Bayes factors, $\log_{10}{\cal B}$ (or $\cal B$), where $0.5<\log_{10}{\cal B}<1$ ($3.2<{\cal B}<10$) is \textit{mild} evidence, $1<\log_{10}{\cal B}<2$ ($10<{\cal B}<100$) is \textit{strong} evidence, and values above this are \textit{decisive}.

Table~\ref{tab:bf} summarizes the evidence for the fiducial model, targeted reductions, and the LVK \bpltwopeak model.
All values are reported relative to the fiducial \cpeaknew model.
The fiducial model is favored over the LVK \bpltwopeak baseline by $\log_{10}\mathcal{B}=7.69$.

Among the targeted reductions, collapsing the two dynamical 1G components into one [D] gives the largest evidence penalty, with $\log_{10}\mathcal{B}=-1.90$.
Sharing mass-ratio distributions within channel type [Q] is also disfavored, with $\log_{10}\mathcal{B}=-1.04$, while sharing the 1G spin-magnitude distribution between isolated and dynamical components [S] gives a smaller penalty, with $\log_{10}\mathcal{B}=-0.56$.
The shared-tilt model [T] is disfavored at a similar level to [Q], but this should be interpreted as evidence against the specific shared isotropic$+$aligned mixture tested here, rather than as a direct measurement of the dynamical tilt distribution.
Thus, the strongest support for the component-level decomposition comes from the mass and mass-ratio structure, while the spin information primarily helps distinguish the hierarchical-labeled component from the 1G components.
Further tests are provided in Appendix~\ref{app:modelcomp}.

\begin{figure*}
    \centering
    \includegraphics[width=\linewidth]{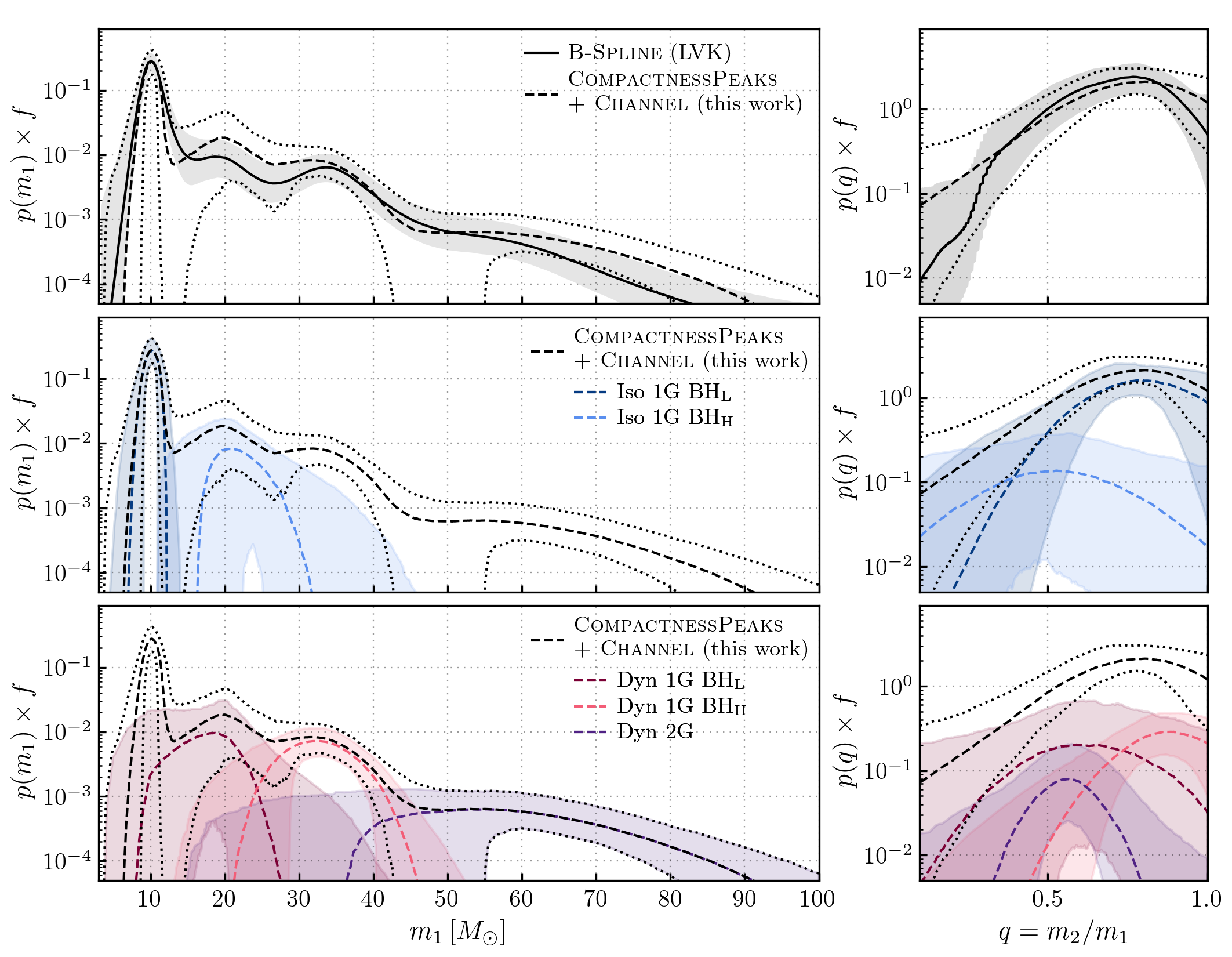}
    \caption{Primary-mass ($m_1$) and mass ratio ($q$) decomposition of the fiducial five-component model, with means and 90\% credible intervals shown.
    The total inferred distribution (mean, dashed line; credible interval, dotted line) is shown in all panels.
    The top row compares the total fiducial model with the GWTC-4.0 \bspline reconstruction for reference.}
    \label{fig:mass}
\end{figure*}

\begin{figure}[t]
    \centering
    \includegraphics[width=\linewidth]{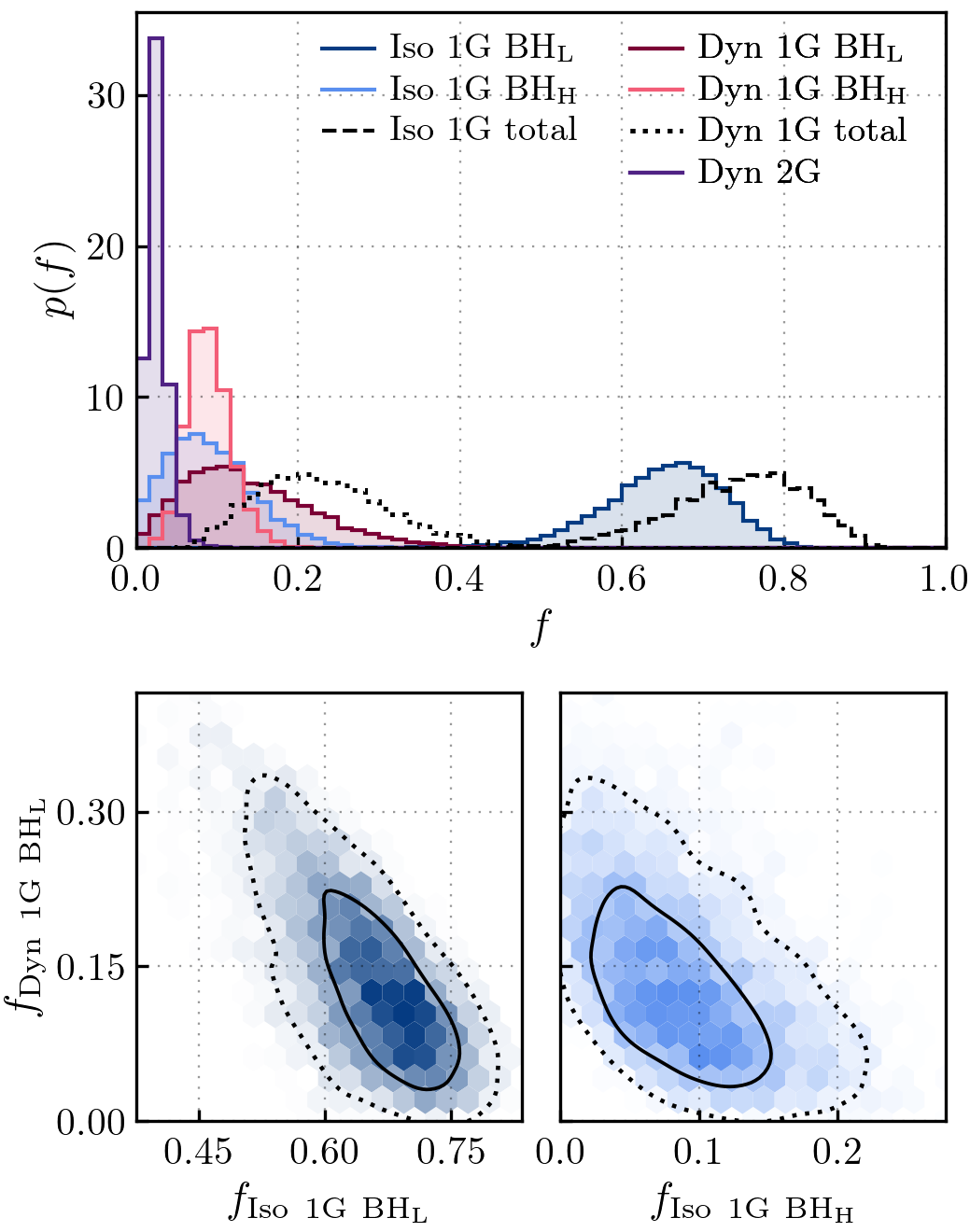}
    \caption{Inferred mixture fractions for the five components of the fiducial \cpeaknew model. \textit{Top}: marginal posteriors for each component fraction $f_X$ (filled histograms), together with the channel-level totals $f_{\rm Iso\,1G} = f_{\IsoL}+f_{\IsoH}$ (dashed) and $f_{\rm Dyn\,1G} = f_{\DynL}+f_{\DynH}$ (dotted). \textit{Bottom}: joint posteriors (50\% and 90\% credible contours) showing the partial degeneracies between $f_{\IsoL}$ and $f_{\DynL}$ (\textit{left}) and between $f_{\IsoH}$ and $f_{\DynL}$ (\textit{right}); the anti-correlations reflect the overlap of the isolated and low-mass dynamical 1G components in the $\sim 10$--$30\,\Msun$ range, where the data do not cleanly isolate their relative contributions.}
    \label{fig:fractions}
\end{figure}

\begin{figure*}[t]
    \centering
    \includegraphics[width=\linewidth]{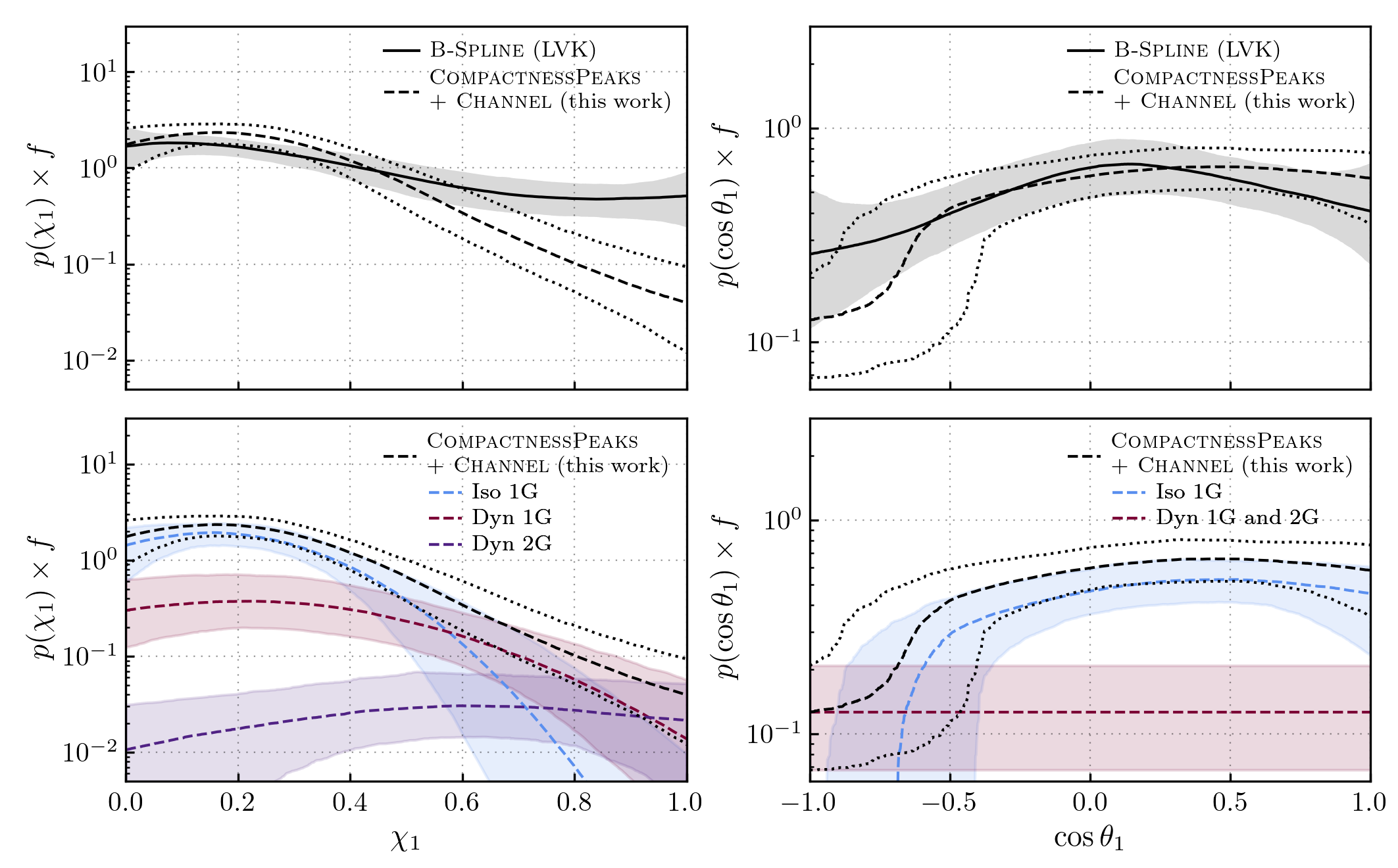}
    \caption{Spin magnitude ($\chi_1$) and spin-tilt ($\cos\theta_1$) decomposition across channels.
    The top row compares the total fiducial model with the GWTC-4.0 \bspline reconstruction for reference; note that the \bspline analysis includes GW231123, which is excluded from our fiducial fit. The bottom row shows the channel-level decomposition.}
    \label{fig:spin}
\end{figure*}

\subsection{Mass structures}

Figure~\ref{fig:mass} shows the inferred primary-mass and mass-ratio population decomposition for the fiducial model, with a comparison to a more data-driven model, the GWTC-4.0 \bspline reconstruction, for reference \citep{GWTC4pop}.
The \bspline reconstruction is shown only as a qualitative reference, since it includes GW231123, whereas our fiducial analysis excludes this event.
The fiducial model captures the main qualitative features of the observed mass spectrum while providing an astrophysically interpretable decomposition into isolated 1G, dynamical 1G, and 2G contributions. Figure~\ref{fig:fractions} shows the inferred component and channel fractions.

\begin{itemize}

    \item \textbf{The $\sim10\,\Msun$ feature.}
    The low-mass isolated component $\IsoL$ is the most sharply localized feature in the fiducial decomposition, with a peak at $\mu_{m_1}^{\IsoL} = \MuIsoLp\,\Msun$, a narrow width, and an upper boundary near $m^{\IsoL}_{\max}=\MmaxIsoLp\,\Msun$.
    Its mass-ratio distribution is moderately narrow and favors near-equal-mass pairings ($\mu_q^{\IsoL}=\PostQMuIsoLp$, $\sigma_q^{\IsoL}=\PostQSigIsoLp$). The inferred mixture fraction is $f_{\IsoL}=\FracIsoLp$.

    \item \textbf{The ``compactness gap'' ($\sim 10$--$15\,\Msun$).}
    The inferred upper edge of $\IsoL$ and lower edge of $\IsoH$ imply a derived gap width
    \begin{equation}
    \Delta m_{\rm gap} = m^{\IsoH}_{\min} - m^{\IsoL}_{\max}.
    \end{equation}
    For the fiducial model, $\Delta m_{\rm gap}=\GapWidth\,\Msun$.
    Because the credible interval is broad and overlaps zero, we do not find clear evidence for an empty component-mass interval.
    In additional tests beyond the fiducial model, the $\IsoH$ component is modeled as a power law with a smooth turn-on.
    In this model the lower edge of $\IsoH$ is better constrained, giving $\Delta m_{\rm gap} = 4.6^{+4.6}_{-4.5}~\Msun$; however, the model is equally preferred to the fiducial model within the typical evidence uncertainty (see Table~\ref{tab:more_bf}).
    The inferred gap width is therefore sensitive to the parameterization of $\IsoH$ (see Appendix~\ref{app:modelcomp}).

    \item \textbf{The intermediate-mass structure ($\sim20\,\Msun$).}
    The $15$--$30\,\Msun$ range is described by overlapping contributions from $\IsoH$ and $\DynL$ (Figures~\ref{fig:mass}, \ref{fig:fractions}), whose mixture fractions are comparable and anti-correlated ($f_{\IsoH}=\FracIsoHp$, $f_{\DynL}=\FracDynLp$). The data do not cleanly separate their relative contributions in this mass range. Their mass-ratio means are correspondingly weakly constrained, and $\IsoH$ in particular is inferred to have the broadest mass-ratio distribution of all five components ($\mu_q^{\IsoH}=\PostQMuIsoHp$, $\sigma_q^{\IsoH}=\PostQSigIsoHp$; $\mu_q^{\DynL}=\PostQMuDynLp$, $\sigma_q^{\DynL}=\PostQSigDynLp$). Removing $\IsoH$ entirely is mildly disfavored ($\mathcal{B}\approx5.2$), indicating some support for structure here but no clear evidence.

    \item \textbf{The $\sim35\,\Msun$ feature.}
    The higher-mass dynamical 1G component ($\DynH$) peaks at $\mu_{m_1}^{\DynH}=\MuDynHp\,\Msun$ with a well-constrained, moderately narrow mass-ratio distribution strongly favoring near-equal-mass systems ($\mu_q^{\DynH}=\PostQMuDynHp$, $\sigma_q^{\DynH}=\PostQSigDynHp$), and a mixture fraction $f_{\DynH}=\FracDynHp$.
    Keeping the two dynamical 1G components separate is strongly favored ($\mathcal{B}\approx80$).

    \item \textbf{The high-mass tail ($>60\,\Msun$).}
    The hierarchical component $\DynTwoG$ dominates the high-mass tail, with a primary-spin distribution shifted to larger values than the 1G components and the narrowest, most tightly constrained mass-ratio distribution of the five components, peaking at a moderately asymmetric value ($\mu_q^{\DynTwoG}=\PostQMuDynTwog$, $\sigma_q^{\DynTwoG}=\PostQSigDynTwog$). Its inferred mixture fraction is small ($f_{\DynTwoG}=\FracDynTwoG$). The component's primary-mass distribution is broad, however, with non-negligible support extending well below the PISN-gap region down to $\sim 25\,\Msun$ (Figure~\ref{fig:mass}); the component is therefore not strictly confined to the high-mass tail.

\end{itemize}

\subsection{Spin distributions}
\label{sec:spin}

Figure~\ref{fig:spin} shows the spin-magnitude and spin-tilt decomposition for the channel-level populations.

The 1G spin magnitudes are inferred to be small in both the isolated and dynamical channels ($\mu_\chi^{\rm Iso}=\MuChiIso$, $\mu_\chi^{\rm Dyn}=\MuChiDyn$), while the hierarchical $\DynTwoG$ component favors substantially larger primary spin magnitudes ($\mu_\chi^{\DynTwoG}=\MuChiDynTwoG$). Fixing this to the spin prediction for remnant spins from numerical relativity ($\mu_\chi^{\DynTwoG}=0.7$) gives a Bayes factor statistically indistinguishable from the fiducial ($\log_{10}\mathcal{B}=-0.16$; Table~\ref{tab:more_bf}), and the freely fitted value $\mu_\chi^{\DynTwoG}=\PostChiMuDynTwog$ agrees with this prediction within the 90\% credible interval.
Allowing separate 1G spin-magnitude distributions [S] is only mildly disfavored ($\log_{10}\mathcal{B}=-0.56$), whereas collapsing all components to a single spin distribution is strongly disfavored ($\log_{10}\mathcal{B}=-2.54$).

The spin-tilt distributions differ by construction across channels: the isolated-channel tilt is modeled as a truncated Gaussian and is inferred to be shifted toward positive $\cos\theta$, with a lower truncation $\cos\theta_{\rm trunc}=\CostTruncIso$, while the dynamical channels are isotropic.
A variation that instead allows every component an independent isotropic--aligned mixture (aligned fraction $\zeta$) is disfavored relative to the fiducial ($\log_{10}\mathcal{B}=-1.14$; Appendix~\ref{app:modelcomp}).
The posteriors of this variant are broad, but generally recover the fiducial assumption with the dynamical components peaking at $\zeta=0$ ($\zeta_{\DynOneG}< 0.83$, $\zeta_{\DynTwoG}<0.8$ at 90\% credibility), consistent with isotropic spins.

\subsection{Redshift evolution}
\label{sec:redshift}

All five components share a single power-law redshift evolution, $\mathcal{R}(z)\propto(1+z)^\kappa$, for which we infer $\kappa=\PostLamb$. This is consistent with the GWTC-4.0 \textsc{Power Law Redshift} result, $\kappa=3.2^{+0.94}_{-1.00}$ \citep{GWTC4pop}, and with the slope of the cosmic star-formation rate density, $\kappa_{\rm SFR}\approx2.7$ \citep{Madau:2014bja}.


\section{Discussion}
\label{sec:discussion}

\subsection{Astrophysical interpretation}
\label{sec:disc_astro}

The fiducial model favors a BBH population in which isolated 1G, dynamical 1G, and hierarchical-merger channels all contribute, broadly consistent with other recent GWTC-4.0 analyses \citep[e.g.][]{Banagiri:2025dmy,Sridhar:2025kvi,Tong:2025xir,Farah:2026jlc,Vijaykumar:2026zjy,Ray:2026uur,Plunkett:2026pxt}.
The added value of the present framework is that the compactness-motivated decomposition of the isolated channel and the further split of the dynamical 1G channel are supported by the data, with component-level mass-ratio information carrying more weight than component-level spin-magnitude structure.

The low-mass end of the spectrum is where the compactness-driven picture holds up best.
The sharply localized $\IsoL$ peak, its turn-off near $\MmaxIsoLp\,\Msun$, its near-equal-mass pairing, and its low, partially aligned spins together match what is expected for black holes formed through stripped-star evolution in close isolated binaries \citep{Schneider:2023mxe,Kalogera:1999tq,Bavera:2020inc,Fuller:2019sxi}, and echo independent identifications of a low-mass isolated-evolution subpopulation \citep{Godfrey:2023oxb}. 
Whether the accompanying dearth above this peak is a genuine ``compactness gap'' remains unresolved: the data localize the compactness-peak edges but not their detailed shapes, and the inferred gap width depends on how the high-mass isolated component is parameterized (Appendix~\ref{app:modelcomp}).
The same low-mass feature is also consistent with a sharper failed-supernova island \citep{Disberg:2023gel,Legred:2026oiz}, which the present analysis does not distinguish from the compactness interpretation.

In our fiducial decomposition, the $\sim35\,\Msun$ feature is dominated by the high-mass dynamical 1G component, whose tightly constrained near-equal-mass distribution is consistent with other analyses \citep{Sridhar:2025kvi,Roy:2025ktr,Ray:2026uur}.
However, an isolated-channel contribution is not excluded.
Stable mass transfer or common-envelope evolution can also populate this mass range \citep{Mandel:2018hfr,vanSon:2022myr,Broekgaarden:2021efa,Li:2025rap}.
Because the high-mass isolated component is broad and overlaps the dynamical population, the present data cannot robustly determine its precise contribution to the $20$--$35\,\Msun$ region.
Structure near $\sim 20\,M_\odot$ was previously highlighted in
parametric GWTC-3 analyses \citep[e.g.][]{Gennari:2025nho}, consistent with the mild support we infer in this work.

The upper edges of the high-mass 1G components may also encode the lower edge of the pair-instability supernova (PISN) gap.
In the fiducial model, the two high-mass 1G components are allowed to have independent upper truncation masses.
The isolated high-mass component turns off at $m_{\max}^{\IsoH}=\PostMOneMaxIsoHp\,\Msun$, while the high-mass dynamical 1G component extends to $m_{\max}^{\DynH}=\PostMOneMaxDynHp\,\Msun$.
This difference should not be over-interpreted as a direct measurement of two distinct PISN edges: both truncation parameters are broad, and the components partially trade support with each other in the $30$--$60\,\Msun$ range.
Rather, the result is consistent with a suppression of the 1G population at high masses, while still allowing channel-dependent flexibility in how that suppression is approached.
As a more direct test, we also consider a variant in which the high-mass isolated component is modeled as a power law with a hard upper cutoff, with this cutoff tied to that of the high-mass dynamical 1G component to represent a common PISN-scale truncation mass.
This shared-cutoff model is statistically indistinguishable from the fiducial ($\log_{10}\mathcal{B}=-0.25$; Table~\ref{tab:more_bf}, Appendix~\ref{app:modelcomp}), and yields $m_{\rm max}^{\rm 1G}=64.3^{+30.3}_{-23.5}\,\Msun$.
The broad credible interval indicates that the data are compatible with a common PISN-scale suppression of the high-mass 1G population without uniquely identifying its location.
Recent work reports evidence for such an edge near $\sim45\,\Msun$ \citep[e.g.][]{Tong:2025wpz,Antonini:2025ilj,Afroz:2025ikg}, although its statistical significance and robustness to modeling assumptions remain debated \citep[e.g.][]{Ray:2025xti,Mould:2026sww}, and its mapping to stellar physics remains affected by uncertain massive-star evolution and nuclear-burning inputs \citep[e.g.][]{Farmer:2019jed,Farmer:2020xne,Renzo:2020rzx,Winch:2024xdt}.

The hierarchical component is the most internally consistent of the higher-mass structures: its elevated primary spin and tightly constrained, moderately asymmetric mass ratio are jointly characteristic of 2G+1G mergers \citep{Gerosa:2017kvu,Rodriguez:2019huv} and difficult to produce through isolated evolution \citep{Zevin:2022bfa}.
This component dominates the high-mass tail; however, the distribution is very broad.
Recent studies have shown that lower-mass hierarchical systems may also contribute to the population at low masses \citep{Tong:2025xir,Farah:2026jlc,Vijaykumar:2026zjy}, along with the discovery of GW241011 and GW241110 as candidate low-mass hierarchical mergers \citep{GW241011_GW241110}.

The shared redshift index, $\kappa=\PostLamb$, is consistent with the cosmic star-formation rate \citep{Madau:2014bja} and with the GWTC-4.0 \textsc{Power Law Redshift} fit \citep{GWTC4pop}.

\subsection{Caveats}
\label{sec:caveats}
Several caveats are worth emphasizing.
First, as noted in Section \ref{sec:model}, the channel labels are astrophysically motivated rather than direct classifications of individual events, and the inferred properties should be interpreted at the population level.
Second, the spin-tilt assumptions should not be over-interpreted.
Recent work \citep{Wolfe:2026meb} cautions that GWTC-4 spin-tilt measurements do not yet provide model-independent evidence for a preferred tilt peak or for a robust correlation between black-hole mass and spin orientation.
This is consistent with our interpretation that the component decomposition is determined primarily by the mass and mass-ratio structure, while the spin information contributes most clearly through the hierarchical-labeled component and remains model dependent.
Although the fiducial model disfavors some more restrictive tilt variants, we do not interpret this as a precise measurement of channel-dependent tilt distributions.
In particular, we assume isotropic tilt distributions for the dynamical components.
This is a sensible first approximation for cluster assembly, but it is not universally appropriate for all dynamical environments.
For example, AGN disks \citep[e.g.][]{Bartos:2016dgn,Stone:2016wzz,McKernan:2019beu} and triple systems \citep[e.g.][]{Stegmann:2025zkb} can produce spin-orbit correlations or misalignments that are not captured by a fixed isotropic prescription.
Third, the present analysis uses a single shared redshift-evolution parameter across all components, although different formation channels may trace different delay-time, metallicity, and cosmic-star-formation histories \citep{Madau:2014bja,Fishbach:2018edt,vanSon:2021zpk}.
Several recent GWTC-4 analyses find evidence that channel contributions vary with redshift \citep[e.g.][]{Banagiri:2025dmy,Farah:2026jlc,Ray:2026uur}, and non-parametric reconstructions suggest that individual mass structures, including the low-mass $\sim 10\,M_\odot$ peak, may follow distinct redshift evolutions \citep{Gennari:2026dfy}.
Resolving component-level redshift evolution is therefore a natural extension of this analysis.
Finally, the present analysis does not distinguish a compactness-driven dearth from a failed-supernova island as explanations for the low-mass feature \citep{Disberg:2023gel,Legred:2026oiz}; see Section \ref{sec:disc_astro}.


\section{Conclusions}
\label{sec:conclusions}

We have presented an analysis probing stellar physics and formation scenarios of the BBH mergers in GWTC-4.0, extending our earlier framework \citep{Galaudage:2024meo} to a five-component mixture with explicit isolated, dynamical 1G, and hierarchical-merger structure. Our main conclusions are as follows:

\begin{enumerate}
    \item The fiducial five-component model is decisively preferred over the LVK \bpltwopeak baseline ($\log_{10}\mathcal{B}=7.69$), with the strongest internal support for keeping the two dynamical 1G components separate.
    Sharing mass-ratio distributions within channel type also worsens the fit, whereas sharing the 1G spin-magnitude distribution has a smaller impact.

    \item The low-mass isolated component is consistent with the compactness-motivated $\sim10\,\Msun$ feature predicted for stripped-star evolution.
    The data localize the approximate positions of the low- and high-mass compactness-peak structures but do not yet robustly resolve the gap between them; the inferred gap width is sensitive to the parameterization of the high-mass isolated component.

    \item The $\sim35\,\Msun$ feature is well described by a high-mass dynamical 1G component, with a possible additional contribution from the isolated channel that remains to be quantified.
    The hierarchical-labeled component is identified jointly by mass, spin, and mass-ratio properties consistent with 2G+1G mergers.

    \item The spin results are qualitatively consistent with partial alignment in the isolated channel and elevated spins in the hierarchical component.
    The tails of the isolated spin-tilt distribution extending to larger misalignment are an area for future study and may be a limitation of the simple ``aligned + isotropic'' construction.
\end{enumerate}

Overall, the current evidence supports a multi-component interpretation of the GWTC-4.0 BBH population and motivates further tests of the compactness-driven isolated-evolution picture against alternative astrophysical scenarios.
Future work with predictions for $\chieff$ and $\chip$ distributions would likely serve as a stronger discriminant of the five components probed in this work.


\begin{acknowledgments}
The author is thankful to Astrid Lamberts, Tristan Bruel, Simon Stevenson, Noah Wolfe, Aditya Vijaykumar, Anarya Ray, Vicky Kalogera and Mike Zevin for their helpful comments and discussions.
Galaudage is supported by CIERA, the Adler Planetarium, and the Brinson Foundation through a CIERA-Adler postdoctoral fellowship.
The author is grateful for computational resources provided by the LIGO Laboratory and supported by National Science Foundation Grants PHY-0757058 and PHY-0823459.
This material is based upon work supported by NSF's LIGO Laboratory which is a major facility fully funded by the National Science Foundation, as well as the Science and Technology Facilities Council (STFC) of the United Kingdom, the Max-Planck-Society (MPS), and the State of Niedersachsen/Germany for support of the construction of Advanced LIGO and construction and operation of the GEO600 detector. Additional support for Advanced LIGO was provided by the Australian Research Council. Virgo is funded, through the European Gravitational Observatory (EGO), by the French Centre National de Recherche Scientifique (CNRS), the Italian Istituto Nazionale di Fisica Nucleare (INFN) and the Dutch Nikhef, with contributions by institutions from Belgium, Germany, Greece, Hungary, Ireland, Japan, Monaco, Poland, Portugal, Spain. KAGRA is supported by Ministry of Education, Culture, Sports, Science and Technology (MEXT), Japan Society for the Promotion of Science (JSPS) in Japan; National Research Foundation (NRF) and Ministry of Science and ICT (MSIT) in Korea; Academia Sinica (AS) and National Science and Technology Council (NSTC) in Taiwan.
\end{acknowledgments}

\section*{Data Availability}
Supplementary material including analysis inputs, posterior samples and additional plots are available at: \href{https://github.com/shanikagalaudage/compactness-peaks-channels}{https://github.com/shanikagalaudage/compactness-peaks-channels}

\software{
GWPopulation \citep{gwpopulation},
\texttt{dynesty} \citep{dynesty},
numpy \citep{numpy},
scipy \citep{scipy},
matplotlib \citep{matplotlib}
}


\appendix
\section{Model variations and feature robustness}
\label{app:modelcomp}

We explore a broader set of restrictions and extensions to the fiducial \cpeaknew model to assess which features are robustly required by the data. Table~\ref{tab:more_bf} reports log Bayes factors and maximum log-likelihood differences relative to the fiducial model.
The upper block contains variations within $|\log_{10}\mathcal{B}|<0.5$ of the fiducial; the lower block contains more strongly disfavored configurations.
Mass-related variations are discussed in Appendix~\ref{app:massdecomp} and spin-related variations in Appendix~\ref{app:spindecomp}.

\begin{table*}
\centering
\begin{tabular}{l c c}
    \hline
    Variation & $\log_{10}\mathcal{B}$ & $\Delta\log_{10}\mathcal{L}_{\rm max}$ \\
    \hline\hline
    Fiducial & $+0.00$ & $+0.00$ \\
    $\mu_\chi^{\DynTwoG}=0.7$ & $-0.16$ & $-0.56$ \\
    $\IsoH = \mathrm{PL}$, shared 1G $m_{\rm max}$ & $-0.25$ & $-0.52$ \\
    $\IsoH = \mathrm{PL}$, $\DynH = \mathrm{BPL}$ & $-0.25$ & $-0.80$ \\
    $\IsoH = \mathrm{PL}$ & $-0.44$ & $-0.50$ \\
    \hline
    $f_{\IsoH}=0$ & $-0.72$ & $-1.05$ \\
    Different $\cos\theta$ per channel, each isotropic$+$aligned mix & $-1.14$ & $-0.34$ \\
    Shared $\cos\theta$ across all, fully isotropic & $-1.41$ & $-1.57$ \\
    Shared $q$ across all channels & $-2.24$ & $-4.27$ \\
    Shared $\chi$ across all channels & $-2.54$ & $-3.52$ \\
    $\cos\theta_{\mathrm{trunc}}=0$ & $-2.74$ & $-1.76$\\
    \hline
\end{tabular}
\caption{
Log Bayes factors and maximum log-likelihood differences for variations of the \cpeaknew model relative to the fiducial configuration.
Negative values indicate that the variation is disfavored.
The upper block contains variations within $|\log_{10}\mathcal{B}|<0.5$ of the fiducial, essentially models that have comparable support from the data; the lower block contains more strongly disfavored configurations.}
\label{tab:more_bf}
\end{table*}

\subsection{Mass-shape variations}
\label{app:massdecomp}

The mass variations in Table~\ref{tab:more_bf} test whether the compactness-motivated structures inferred in the fiducial model depend on the assumed shape of the high-mass isolated and dynamical components.
First, replacing the truncated-Gaussian $\IsoH$ component with a power law and a smooth low-mass turn-on is statistically comparable to the fiducial model, with $\log_{10}\mathcal{B}=-0.44$.
In this parameterization, $\IsoH$ turns on over a window from $m_{\min}^{\IsoH}$ to $m_{\min}^{\IsoH}+\delta_m^{\IsoH}$.
The smoothing width $\delta_m^{\IsoH}$ is weakly constrained, but the turn-on location $m_{\min}^{\IsoH}$ is more localized than the lower truncation of the fiducial truncated-Gaussian $\IsoH$ component.
This leads to a more clearly gap-like inferred shape, with $\Delta m_{\rm gap}=4.6^{+4.6}_{-4.5}\,\Msun$.
However, because this model is not preferred over the fiducial model, we do not interpret this as stronger evidence for a resolved compactness gap.
Instead, it shows that the inferred gap width is sensitive to the adopted shape of $\IsoH$.
Second, tying the high-mass cutoff of the power-law $\IsoH$ component to the upper cutoff of $\DynH$ gives $\log_{10}\mathcal{B}=-0.25$, again statistically indistinguishable from the fiducial model.
This indicates that the data do not require separate high-mass cutoffs for the isolated and dynamical 1G components.
The common cutoff in this model is broad, $m_{\rm max}^{\rm 1G}=64.3^{+30.3}_{-23.5}\,\Msun$, so it should be interpreted as a consistency check on a shared PISN-scale suppression rather than as a precise measurement of the lower edge of the PISN gap.
Third, replacing the high-mass dynamical 1G Gaussian with a broken power law gives $\log_{10}\mathcal{B}=-0.25$.
Thus, the data prefer support near the $\sim35\,\Msun$ region, but the present catalog does not uniquely determine whether this structure is best described as a localized peak or as part of a broader high-mass continuum.
Finally, suppressing $\IsoH$ entirely by setting $f_{\IsoH}=0$ is mildly disfavored, with $\log_{10}\mathcal{B}=-0.72$.
This gives mild, model-dependent support for additional structure in the $15$--$30\,\Msun$ range, but does not establish a distinct $\sim20\,\Msun$ feature.

\subsection{Spin-distribution variations}
\label{app:spindecomp}

Fixing $\mu_\chi^{\DynTwoG}=0.7$ is statistically indistinguishable from the fiducial fit, with $\log_{10}\mathcal{B}=-0.16$.
This supports the interpretation that the hierarchical-labeled component is compatible with the expected remnant spin of comparable-mass black-hole mergers, while also showing that the data do not require the 2G spin location to be freely fitted.

The fiducial model treats spin tilts asymmetrically: the isolated channel has a preferentially aligned truncated Gaussian, while the dynamical channels are fixed to be isotropic.
We test this assumption with several alternatives.
Allowing each channel to have its own isotropic$+$aligned mixture gives $\log_{10}\mathcal{B}=-1.14$.
Forcing all channels to share a single isotropic$+$aligned mixture gives $\log_{10}\mathcal{B}=-1.16$.
Forcing all channels to be isotropic gives $\log_{10}\mathcal{B}=-1.41$.
Within this model family, the data prefer not to collapse all tilt information into a single shared prescription.
However, the posteriors on the dynamical-channel aligned fractions are broad, so this should not be interpreted as a precise measurement of dynamical isotropy.

We also test a more restrictive isolated-channel tilt model in which $\cos\theta_{\rm trunc}=0$, so that the isolated spin-tilt distribution has support only on prograde configurations.
This choice is motivated by the expectation that isolated binary evolution should preferentially produce small spin--orbit misalignments, although natal kicks, mass transfer, and uncertain angular-momentum transport can broaden the distribution.
This variant is strongly disfavored, with $\log_{10}\mathcal{B}=-2.74$ and $\Delta\log_{10}\mathcal{L}_{\max}=-1.76$.
The decrease in maximum likelihood shows that the evidence penalty is not purely an Occam effect: imposing a hard prograde-only lower bound also worsens the best possible fit.
We therefore interpret this result as evidence that this simple prograde-only prescription is too restrictive within the present five-component decomposition, rather than as evidence that isolated binaries generically require retrograde spins.
In particular, the fiducial isolated tilt model should be viewed as an effective population-level description that allows partial misalignment, not as a direct measurement of the spin-tilt distribution of a pure isolated formation channel.

The strongest spin-magnitude evidence comes from distinguishing the hierarchical-labeled component from the 1G components.
Sharing the 1G spin-magnitude distribution between isolated and dynamical components is only mildly disfavored, whereas sharing the spin distribution across all components is more strongly disfavored, with $\log_{10}\mathcal{B}=-2.54$.
Thus, the data require spin flexibility mainly to accommodate the high-spin hierarchical-labeled component, while the evidence for distinct isolated and dynamical 1G spin-magnitude distributions is weaker.
Taken together, these tests suggest that the present data are more sensitive to the existence of a high-spin hierarchical component than to the detailed separation of the isolated and dynamical 1G spin distributions.
At the same time, overly restrictive tilt prescriptions can degrade the fit, emphasizing that the channel labels should be interpreted as population-level components rather than clean event-by-event formation assignments.

\section{Population model hyperparameters and hyperpriors}
\label{app:priors}

Table~\ref{tab:compactnesspeaks_priors} summarizes the hyperpriors used for the fiducial \cpeaknew model. The mixture fractions $f_X$ are drawn from a symmetric Dirichlet distribution that is uninformative over the simplex, allowing each of the five components to vary freely between $0$ and $1$ subject to $\sum_X f_X = 1$.

The $m_1$ priors are designed to reflect the astrophysical motivation of each component while remaining broad enough to allow the data to drive the inference. The isolated 1G compactness components ($\IsoL$ and $\IsoH$) are restricted to mass ranges centered on the predicted locations of the stripped-star compactness peaks of \citet{Schneider:2023mxe}: $\mu_{m_1}^{\IsoL}\in(8,12)\,\Msun$ for the low-mass peak and $\mu_{m_1}^{\IsoH}\in(12,60)\,\Msun$ for the high-mass peak.
The truncation bounds $m_{\max}^{\IsoL}\in(10,15)\,\Msun$ and $m_{\min}^{\IsoH}\in(10,25)\,\Msun$ allow the two isolated components to overlap, touch, or separate.
The derived compactness-gap width, $\Delta m_{\rm gap}=m_{\min}^{\IsoH}-m_{\max}^{\IsoL}$, therefore ranges over both positive and negative values under the prior.
This prior structure is designed to test whether the data favor separation between the two compactness-motivated isolated components, rather than imposing a gap a priori.
The hierarchical 2G component ($\DynTwoG$) is given a broad prior that extends into the predicted PISN-gap region, accommodating a high-mass remnant population.

The mass-ratio distributions use independent truncated Gaussians with weakly informative priors on the mean and width. The spin-magnitude priors reflect channel-specific astrophysical expectations: the isolated 1G mean is restricted to $\mu_\chi^{\rm Iso}\in(0,0.5)$, consistent with predictions of slowly rotating BHs from stellar collapse \citep{Fuller:2019sxi}, while the dynamical 1G and 2G means are given broader and higher ranges, with $\mu_\chi^{\DynTwoG}\in(0.4,1)$ centered on the numerical-relativity prediction of $\chi\approx0.69$ for remnants of comparable-mass mergers \citep{Lousto:2009mf,Hofmann:2016yih}. The isolated spin-tilt distribution is parameterized as a truncated Gaussian whose lower truncation $\cos\theta_{\rm trunc}\in(-1,0)$ allows for partial alignment from tidal interactions in close binaries; dynamical channels have isotropic tilts by construction. A single power-law redshift evolution with index $\kappa$ is shared across all five components.

\begin{figure*}
    \includegraphics[width=\linewidth]{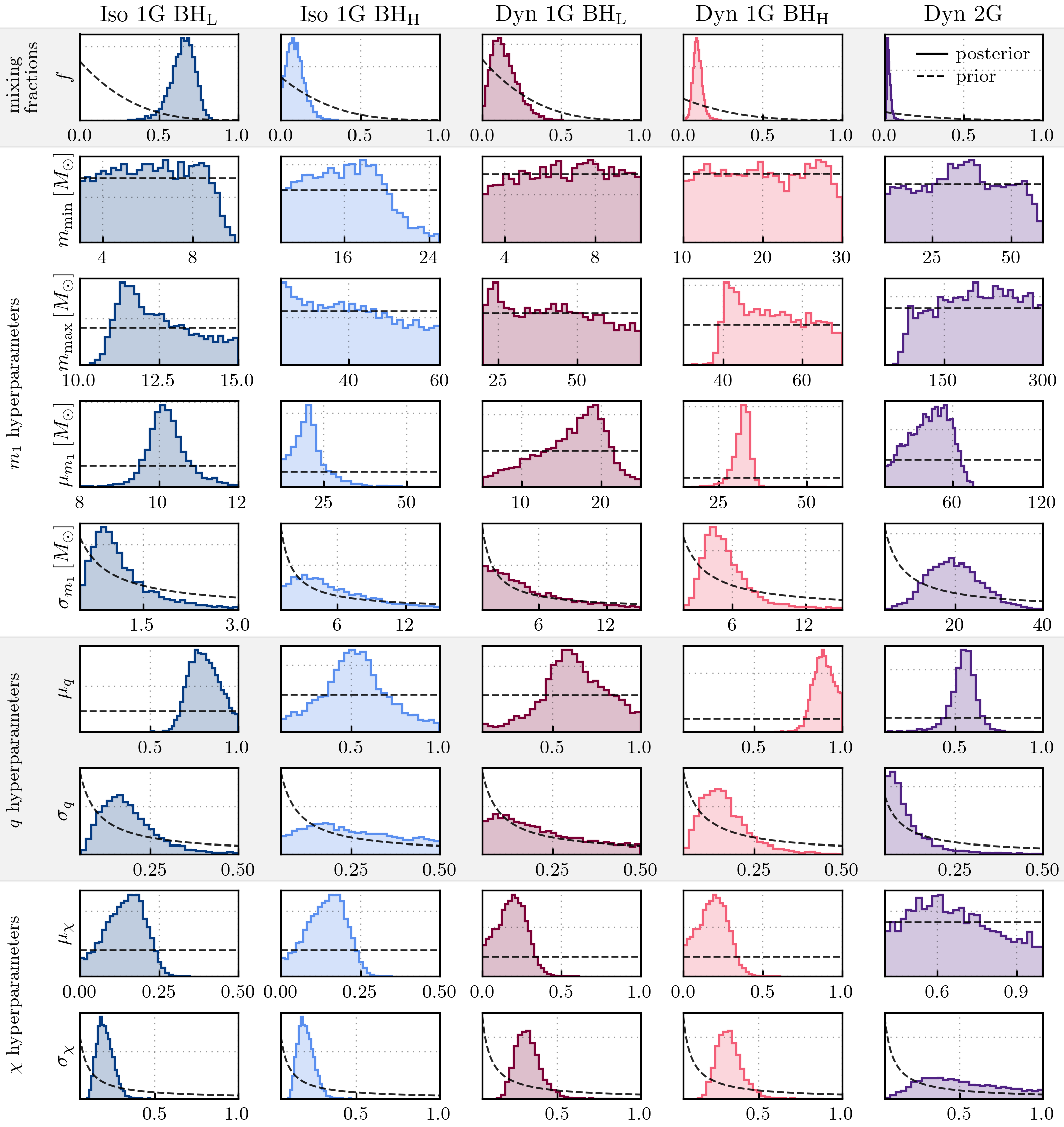}
    \caption{Posterior distributions for the hyperparameters of the fiducial \cpeaknew model (filled), compared to the corresponding priors (dashed). Columns correspond to the five components ($\IsoL$, $\IsoH$, $\DynL$, $\DynH$, $\DynTwoG$); rows show, from top to bottom: mixture fractions, $m_1$ truncation bounds, $m_1$ Gaussian means and widths, $q$ Gaussian means and widths, and $\chi$ Gaussian means and widths. The posteriors are most informative for the $\IsoL$ peak ($\mu_{m_1}^{\IsoL}=\MuIsoLp\,\Msun$), the $\DynH$ peak location and mass-ratio mean ($\mu_{m_1}^{\DynH}=\MuDynHp\,\Msun$, $\mu_q^{\DynH}=\PostQMuDynHp$), and the channel-level mixture fractions; several truncation parameters and the broader $\IsoH$ and $\DynL$ components remain partially prior-dominated.}
    \label{fig:hyperparams_priors}
\end{figure*}

Figure~\ref{fig:hyperparams_priors} provides a prior--posterior comparison for the fiducial hyperparameters.
The most informative posteriors are those associated with the $\IsoL$ peak location and upper edge, the $\DynH$ peak location and mass-ratio distribution, and the component fractions.
By contrast, several width and truncation parameters remain close to their priors, especially for the broader $\IsoH$ and $\DynL$ components.
This motivates our interpretation that the current data localize the main mass structures more robustly than they determine the detailed component shapes.

\newcommand{\ptparam}[1]{\parbox[t]{2.7cm}{\raggedright #1}}
\newcommand{\ptdesc}[1]{\parbox[t]{11.4cm}{\raggedright #1}}
\newcommand{\ptprior}[1]{\parbox[t]{2.6cm}{\raggedright #1}}

\startlongtable
\begin{deluxetable*}{lll}
\tablecaption{Summary of hyperpriors used in the \cpeaknew\ population model for the primary mass ($m_1$), mass ratio ($q$), spin magnitude ($\chi$), spin orientation ($\cos\theta$) and redshift ($z$) distributions.
\label{tab:compactnesspeaks_priors}}
\tablehead{
  \colhead{\textbf{Parameter}} &
  \colhead{\textbf{Description}} &
  \colhead{\textbf{Prior}}
}
\startdata
\hline
\ptparam{$f_X$} &
\ptdesc{Mixture fraction of each mass component, with $\sum_X f_X = 1$.} &
\ptprior{$\mathrm{Dirichlet}(1,1,1,1,1)$} \\
\ptparam{$m_{\min}^{\IsoL}$ [$\Msun$]} &
\ptdesc{Minimum $m_1$ of low-mass isolated 1G component.} &
\ptprior{$\mathcal{U}(3,10)$} \\
\ptparam{$m_{\min}^{\DynL}$ [$\Msun$]} &
\ptdesc{Minimum $m_1$ of low-mass dynamical 1G component.} &
\ptprior{$\mathcal{U}(3,10)$} \\
\ptparam{$m_{\max}^{\IsoL}$ [$\Msun$]} &
\ptdesc{Maximum $m_1$ of isolated low-mass 1G component.} &
\ptprior{$\mathcal{U}(10,15)$} \\
\ptparam{$m_{\min}^{\IsoH}$ [$\Msun$]} &
\ptdesc{Minimum $m_1$ of isolated high-mass 1G component.} &
\ptprior{$\mathcal{U}(10,25)$} \\
\ptparam{$m_{\max}^{\IsoH}$ [$\Msun$]} &
\ptdesc{Maximum $m_1$ of isolated high-mass 1G component.} &
\ptprior{$\mathcal{U}(25,60)$} \\
\ptparam{$m_{\max}^{\DynL}$ [$\Msun$]} &
\ptdesc{Maximum $m_1$ of dynamical low-mass 1G component.} &
\ptprior{$\mathcal{U}(20,70)$} \\
\ptparam{$m_{\min}^{\DynH}$ [$\Msun$]} &
\ptdesc{Minimum $m_1$ of dynamical high-mass 1G component.} &
\ptprior{$\mathcal{U}(10,30)$} \\
\ptparam{$m_{\max}^{\DynH}$ [$\Msun$]} &
\ptdesc{Maximum $m_1$ of dynamical high-mass 1G component.} &
\ptprior{$\mathcal{U}(30,70)$} \\
\ptparam{$m_{\min}^{\DynTwoG}$ [$\Msun$]} &
\ptdesc{Minimum $m_1$ of dynamical 2G component.} &
\ptprior{$\mathcal{U}(10,60)$} \\
\ptparam{$m_{\max}^{\DynTwoG}$ [$\Msun$]} &
\ptdesc{Maximum $m_1$ of dynamical 2G component.} &
\ptprior{$\mathcal{U}(60,300)$} \\
\ptparam{$\mu_{m_1}^{\IsoL}$ [$\Msun$]} &
\ptdesc{Mean of Gaussian $m_1$ for the isolated low-mass 1G component.} &
\ptprior{$\mathcal{U}(8,12)$} \\
\ptparam{$\sigma_{m_1}^{\IsoL}$ [$\Msun$]} &
\ptdesc{Width of Gaussian $m_1$ for the isolated low-mass 1G component.} &
\ptprior{$\mathrm{Log}\mathcal{U}(0.5,3)$} \\
\ptparam{$\mu_{m_1}^{\IsoH}$ [$\Msun$]} &
\ptdesc{Mean of Gaussian $m_1$ for the isolated high-mass 1G component.} &
\ptprior{$\mathcal{U}(12,60)$} \\
\ptparam{$\mu_{m_1}^{\DynL}$ [$\Msun$]} &
\ptdesc{Mean of Gaussian $m_1$ for the dynamical low-mass 1G component.} &
\ptprior{$\mathcal{U}(5,25)$} \\
\ptparam{$\sigma_{m_1}^{\IsoH}$ [$\Msun$]} &
\ptdesc{Width of Gaussian $m_1$ for the isolated high-mass 1G component.} &
\ptprior{$\mathrm{Log}\mathcal{U}(1,15)$} \\
\ptparam{$\sigma_{m_1}^{\DynL}$ [$\Msun$]} &
\ptdesc{Width of Gaussian $m_1$ for the dynamical low-mass 1G component.} &
\ptprior{$\mathrm{Log}\mathcal{U}(1,15)$} \\
\ptparam{$\mu_{m_1}^{\DynH}$ [$\Msun$]} &
\ptdesc{Mean of Gaussian $m_1$ for the dynamical high-mass 1G component.} &
\ptprior{$\mathcal{U}(15,60)$} \\
\ptparam{$\sigma_{m_1}^{\DynH}$ [$\Msun$]} &
\ptdesc{Width of Gaussian $m_1$ for the dynamical high-mass 1G component.} &
\ptprior{$\mathrm{Log}\mathcal{U}(2,15)$} \\
\ptparam{$\mu_{m_1}^{\DynTwoG}$ [$\Msun$]} &
\ptdesc{Mean of Gaussian $m_1$ for the dynamical 2G component.} &
\ptprior{$\mathcal{U}(15,120)$} \\
\ptparam{$\sigma_{m_1}^{\DynTwoG}$ [$\Msun$]} &
\ptdesc{Width of Gaussian $m_1$ for the dynamical 2G component.} &
\ptprior{$\mathrm{Log}\mathcal{U}(4,40)$} \\
\hline
\ptparam{$\mu_q^X$} &
\ptdesc{Mean of Gaussian $q$ for each component.} &
\ptprior{$\mathcal{U}(0.1,1)$} \\
\ptparam{$\sigma_q^X$} &
\ptdesc{Width of Gaussian $q$ for each component.} &
\ptprior{$\mathrm{Log}\mathcal{U}(0.05,0.5)$} \\
\hline
\ptparam{$\mu_\chi^{\mathrm{Iso}}$} &
\ptdesc{Mean of $\chi$ shared by the isolated 1G components.} &
\ptprior{$\mathcal{U}(0,0.5)$} \\
\ptparam{$\mu_\chi^{\mathrm{Dyn}}$} &
\ptdesc{Mean of $\chi$ shared by the dynamical 1G components.} &
\ptprior{$\mathcal{U}(0,1)$} \\
\ptparam{$\mu_\chi^{\DynTwoG}$} &
\ptdesc{Mean of $\chi$ for the primary black hole in the dynamical 2G component.} &
\ptprior{$\mathcal{U}(0.4,1)$} \\
\ptparam{$\sigma_\chi^Y$} &
\ptdesc{Widths of $\chi$ for the isolated 1G, dynamical 1G, and dynamical 2G components.} &
\ptprior{$\mathrm{Log}\mathcal{U}(0.05,1)$} \\
\ptparam{$a_{\max}$} &
\ptdesc{Maximum allowed dimensionless spin magnitude.} &
\ptprior{$1$} \\
\hline
\ptparam{$f_{\mathrm{aligned}}$} &
\ptdesc{Fraction of isolated binaries in the preferentially aligned spin-tilt subpopulation.} &
\ptprior{$1$} \\
\ptparam{$\mu_{\cos t}$} &
\ptdesc{Mean of isolated $\cos\theta$.} &
\ptprior{$\mathcal{U}(0,1)$} \\
\ptparam{$\sigma_{\cos t}$} &
\ptdesc{Width of isolated $\cos\theta$.} &
\ptprior{$\mathrm{Log}\mathcal{U}(0.01,2)$} \\
\ptparam{$\cos\theta_{\mathrm{trunc}}$} &
\ptdesc{Lower truncation point for the isolated $\cos\theta$.} &
\ptprior{$\mathcal{U}(-1,0)$} \\
\hline
\ptparam{$\kappa$} &
\ptdesc{Power-law index of redshift evolution of merger rate.} &
\ptprior{$\mathcal{U}(-10,10)$} \\
\enddata
\tablecomments{
$\mathcal{U}(a,b)$ denotes a uniform prior over the interval $(a,b)$, and
$\mathrm{Log}\mathcal{U}(a,b)$ denotes a log-uniform prior over the interval $(a,b)$. $X\in\{\IsoL,\IsoH,\DynL,\DynH,\DynTwoG\}$, $Y\in\{\mathrm{Iso},\DynOneG,\DynTwoG\}$
}
\end{deluxetable*}

\bibliography{main}{}
\bibliographystyle{aasjournalv7}

\end{document}